\documentclass{article}
\usepackage[a4paper, portrait, margin=1in]{geometry}
\usepackage{setspace}
\usepackage[utf8]{inputenc}
\usepackage[english]{babel}
\usepackage{apacite}
\usepackage{subfig}
\usepackage[english]{babel}
\usepackage{amsmath}
\usepackage{graphicx}
\usepackage[colorinlistoftodos]{todonotes}
\usepackage[left]{lineno}
\usepackage{diagbox}
\usepackage{indentfirst}
\usepackage{adjustbox}
\usepackage{xcolor}
\usepackage{multirow}
\usepackage{lipsum}
\usepackage{sectsty}
\sectionfont{\bfseries\Large\raggedright}
\usepackage{apacite}

\begin{document}

\raggedbottom

\hbadness=99999

\noindent \large \textbf{CHAPTER 9}

\noindent\large \textbf{EMOTIONAL DESIGN}\\

\noindent\textbf{Feng Zhou, Assistant Professor}

\noindent\emph{Department of Industrial and Manufacturing Systems Engineering\\
University of Michigan-Dearborn}\\

\noindent\textbf{Yangjian Ji, Professor}\\
\emph{School of Mechanical Engineering\\
Zhejiang University}\\

\noindent\textbf{Roger Jianxin Jiao, Associate Professor}\\
\emph{The George W. Woodruff School of Mechanical Engineering\\
Georgia Institute of Technology}\\

\noindent\emph{To Appear in the 5th edition of the Handbook of Human Factors and Ergonomics}

\newpage
\tableofcontents
\newpage

\section{INTRODUCTION}
\subsection{What is Emotion}

The concept of emotion is closely related to affect, which is an encompassing term and it consists of emotions, feelings, moods, and evaluations \cite{simon1982affect}. The most important concept is probably emotion. Nevertheless, in psychology, the theories about emotion consider it a ‘very confused and confusing field of study’ \cite{ortony1988cognitive} and thus there is no consensus on a definition. Various factors are associated with emotions, including subjective factors, environmental factors, and neural and hormonal processes. In this chapter, we make use of the summary of emotion provided by Kleinginna and Kleinginna \citeyear{kleinginna1981categorized}, which incorporates the key elements of definitions in psychology as follows: 

\noindent (1) Emotions give rise to affective experience, such as pleasure or displeasure.\\
(2) Emotions stimulate us to generate cognitive explanations – to attribute the cause to ourselves or to the environment, for example.\\
(3) Emotions trigger a variety of internal adjustments in the autonomic nervous system, such
as an increased heart rate and a decreased skin conductance response.\\
(4) Emotions elicit behaviors that are often, but not always, expressive (laughing or crying), goal-directed (approaching or avoiding), and adaptive (removal of a potential threat).

Feelings can be used to describe physical sensation of touch through either experience or perception, and are subjective representations of emotions, which can be consciously felt \cite{davidson2009handbook}. Thus, they are often used as self-reported measures for emotions in the literature \cite<e.g.,>{zhou2011affect}. 

Moods are associated with affective states with a longer duration  \cite{picard1997affective}. They can last for hours, days, or even longer without an attributed object. Emotions are often short-lived, but when an emotion, thought, or action, is repeatedly activated, it can result in a mood \cite{russell2003core}. For instance, a negative mood can be produced by repeated negative emotions, thoughts, actions or induced by drugs or medication \cite{picard1997affective}.

Subjective evaluation is often defined as a valenced affective response that can assess an object or a situation with positive or negative opinions, views, or reactions \cite{simon1982affect}.  

Russell \citeyear{russell2003core} used \emph{core affect} to describe all the emotionally charged events, including emotion, mood, and evaluation. It has two important dimensions, i.e., valence (pleasure-displeasure) and arousal (sleepy-activated). Compared to the discrete emotion models, such as basic emotions proposed by Paul Ekman \citeyear{ekman1992argument}, who argued that there were six basic emotions (i.e., anger, disgust, fear, happiness, sadness, surprise) and that they could be recognized by facial expressions across different cultures, Russell \citeyear{russell1989affect} argued that valence and arousal were two important dimensions in the continuous emotion model. Individual emotions can be specified with these two dimensions. For example, excitement is characterized by positive valence and high arousal while sad is characterized by negative valence and low arousal. 

\subsection{Emotion in Human Factors and Ergonomics}
Traditional human factors and ergonomics (HFE) researchers mainly addressed the physical and cognitive aspects of the human to prevent frustration, pain, stress, fatigue, overload, injury, and death in the design, development, and deployment of products and systems \cite{wickens1998introduction,hancock2005hedonomics}. Since the 1990s, researchers in HFE have started to advocate positive experience of the human, including flow experience \cite{csikszentmihalyi1990flow} and hedonomics \cite{helander2003hedonomics,hancock2005hedonomics}. Contrary to traditional HFE which prevented negative aspects of human experience, positive psychology advocated positive aspects of human experience, such as happiness, well-being, and positivity \cite{csikszentmihalyi2000positive}. This notion started to influence HFE and one good example is the concept of the flow experience, in which one is so intensely absorbed and immersed in the task that it results in positive emotions, exploratory behavior, and behavioral perceived control \cite{csikszentmihalyi1990flow} during the human-product interaction. This can only happen when the task difficulty level matches the user's skill level with a clear set of goals and immediate feedback. The view of positive psychology further influenced pioneer researchers  in HFE for pursuing hedonomics \cite{helander2003hedonomics,hancock2005hedonomics,helander2005underlying}. It promotes pleasurable experience and individuation in the process of human-product interaction. Pleasurable experience goes beyond safety, reliability, and usability to include joy, fun, and positive experience resulted from users' appraisal, perception, and interaction with the product while individuation emphasizes customization and personalization of the tools for individuals to optimize efficiency and pleasure \cite{hancock2005hedonomics}. Recently, hedonomics has been proposed to reach its fullest potential to collective goals in organizational and social contexts, such as the workplace \cite{oron2017ergonomics}.    

In addition, organizations, conferences, and special issues related to emotion and design in HFE have also been burgeoning. In 1999, the Design and Emotion Society was built \cite{desmet1999love} with the First International Conference on Design and Emotion held in Delft, The Netherlands. Since then, it has been held bi-annually, where researchers and industry practitioners and leaders interact with each other in the domain of design and emotion. At the 10-th anniversary of the International Conference on Design and Emotion, a special issue was created in the \emph{International Journal of Design} to synthesize different design and emotion related studies \cite{desmet2009special}. In addition, the International Conference on Kansei Engineering and Emotion Research was created in 2007 and held bi-annually to invite related researchers and industrial practitioners and leaders exchange knowledge in emotion and Kansei research \cite{nagamachi1995kansei} in product design and development. Both the International Conference on Kansei Engineering and the International Conference on Affective and Pleasurable Design are affiliated with the International Conference on Applied Human Factors and Ergonomics series. Emotion related topics on design also frequently appear in the ACM CHI Conference on Human Factors in Computing Systems, which is the premier international conference of Human-Computer Interaction, and recent examples include \cite{altarriba2020technology,dmitrenko2020caroma}.

\section{CONNECTING EMOTION TO DESIGN}

\subsection{Emotional Associations} 
Core affect is object-free without directing anything, i.e., no emotional associations, whereas affective quality related to or belonged to the product has the ability to cause a change in core affect during the human-product interaction process so that it can be attributed to the product to create emotional associations \cite{russell2003core,zhou2011fundamentals}. Note core affect is within the user, but affective quality lies in the product. Similar to core affect, affective quality can also be described with valence and arousal as a dimensional construct. Valence, as the intrinsic pleasure or displeasure, of a product feature often governs the fundamental user responses or reactions in the interaction process, i.e., likes and attraction, which encourage approach, versus dislikes or aversion, which lead to withdrawal and avoidance \cite{bradley2001emotion,zhou2011fundamentals}. Despite the distinct personalities, emotional baggage, and unique dispositions, there are common psychology principles that are common to all humans that we can use to build emotional associations \cite{walter2011designing}, such as the baby face bias, the golden ratio rule, and the Gestalt principles. For example, designers can make use of the baby face bias to motivate users and high baby schema infants were considered as more cute and elicited stronger motivation for caretaking than low baby schema infants \cite{glocker2009baby}, the golden ratio rule is widely applied in website design, such as Twitter \cite{walter2011designing}, and Gestalt principles of perceptual organization can make a design coherent and orderly and, therefore, pleasant to look at \cite{desmet2007framework,schifferstein2011product}.

Arousal also influences the resulted emotional responses to human-product interaction. It can be defined as a psychological and physiological level of awakeness and it can influence a person's sensory alertness, mobility, and readiness to respond \cite{kubovy1999pleasures}. Studies showed that there was an optimal level of arousal for individual task performance, i.e., the inverted-U shape Yerkes-Dodson law \cite{yerkes1908relation}. For example, a state of high vigilance is still required in human-automation interaction in conditional automated driving for the driver to be ready for takeover transitions \cite{ayoub2019manual,zhou2020driver,du2020examining}. In other interactive applications and areas, including training, learning, and gaming, an optimal level of arousal is also important in order to maintain or prevent particular alertness for optimal performance and positive emotions \cite{zhou2011affect,zhou2014emotion,zhou2017affective}, which can be similar to the flow experience in the human-product interaction process.

\subsection{Factors Influencing Emotional Experience}
We exam the factors that influence emotional experience using the appraisal theory \cite{ortony1990cognitive, ellsworth2003appraisal,clore2013psychological}, in which human users assess stimuli with regard to their perceived significance considering their goals and needs comparing with their coping capabilities with corresponding consequences and the compatibility of the actions with perceived social norms and self-ideals. Under such a framework, we categorize the factors into human needs, product quality, and ambient factors as well as their dynamic relationships, which was termed as human-product-ambience interaction \cite{zhou2011fundamentals,zhou2013affective}.

According to the motivational theory in psychology \cite{maslow1987maslow}, human needs comprise a five-tier hierarchy, and from the bottom upwards, they are physiological, safety, love and belonging, esteem, and self-actualization. Correspondingly, for product design, human needs can be divided into a similar hierarchy of needs, including functional, reliable, usable, pleasurable, and individuation \cite{hancock2005hedonomics,walter2011designing}. With regard to emotional design, the higher level of user needs that go beyond the instrumental ones \cite{hassenzahl2006user}, i.e., affective needs, including pleasurable and individuation, are defined in a broader perspective to focus on emotional responses and aspirations \cite{jiao2007analytical}, and are deeply implanted in the lower levels of basic needs to minimize pain and maximize pleasure, both psychologically and physically. The strength of such pain or pleasure is built on the user's appraisal process and ensuing results. During the interaction process between the human user and the product, the user evaluates whether the tasks involved are facilitating (affective) needs fulfillment. If so, positive emotional responses can be elicited. 

As mentioned earlier, good affective quality related to or within the product can greatly satisfy affective needs by attributing positive emotional responses to the product (features). For example, if an automated school bus is able to assure safety in transporting children, the parents will have no anxiety or worry, but rather trust and ease \cite{ayoub2020otto}. Moreover, from affective computing's point of view, smart products equipped with emotion sensing capabilities may help frustrated users and prevent other negative emotions (e.g., anger in driving) by designed interventions \cite{picard2002computers}. For example, in education and learning, many researchers made use of the emotional lens to prevent negative emotional responses, optimize learning performance, and advocate positive emotional outcomes \cite{yadegaridehkordi2019affective}.

Consistent with the flow experience \cite{csikszentmihalyi1990flow,csikszentmihalyi2000positive}, the appraisal theory also considers users' ability to deal with the tasks in the human-product interaction process by reaching, modifying, postponing, or giving up goals or needs to modulate their emotional responses \cite{ellsworth2003appraisal}. When one's coping capabilities match the task challenge levels (dynamically), one is able to sustain the flow experience continuously. Examples in HFE often compare novice users with experienced users, old users with young users, and male users with female users, etc. in evaluating product performance, usability, and affective quality. For instance, ordinary use cases were compared with extraordinary use cases in order to elicit latent customer needs that could delight customers unexpectedly \cite{zhou2015latent}. For another example, trust in automated vehicles consisted of multiple interacting variables, including the age of the drivers, risks, and reliability of the vehicle and younger drivers  reduced their trust significantly more than older drivers when there was automation failures \cite{rovira2019looking}.

Other particular ambient factors that can potentially influence one's emotional responses include environmental settings and cultural differences \cite{zhou2011fundamentals,zhou2013affective}. These factors can be considered as moderator variables that can either improve or weaken the relationships between the user factors and product factors. The environmental settings are factors that influence where the product will be used and how the product will be used in combination with other products. These factors affect users' perception of product value and assessment. For example, the interior setting of a plane, including the humidity level, the noise level, the lighting, the interior color and pattern, can significantly influence a passenger's flying experience \cite{zhou2014prospect}.
For another example, a Kindle device is supposed to be used in various environments and places, and the designer need to consider whether it is sensitive to various environmental settings (e.g., light conditions, parental control for kids) \cite{zhou2015latent}. In addition, the sequence effect states that a product is positively evaluated in isolation, but can be eventually not used or possessed due to its unfitness with other products that are previously purchased, including furniture, computer hardware and software, and appliances \cite{bloch1995seeking}. For instance, when Microsoft rolled out the Windows Vista operating system, its compatibility issues (e.g., the Aero interface) caused negative emotional responses \cite{livingston2007windows}. 

Cultural factors can also influence users' perception on products due to the fact that humans are socially living  species. Typical examples include aesthetic stereotypes, national shapes and colors, social rules and norms, historical beliefs, customs, practices, and so on \cite{qin2019impact}. For example, participants from countries with individualistic cultures (e.g., United States, Canada, Germany, and United Kingdom) liked angular patterns while those from 
countries with collective cultures (e.g., Japan, South Korea, and Hong Kong) preferred round patterns. Not only across cultures, designers should take cultural differences into the design process, but also across inter-generations within one culture. For example, how young people perceive traditional cultural design can significantly influence their emotional responses to and attitudes towards cultural product
design \cite{chai2015relative}. 

\subsection{Models and Methods Related to Emotional Design}

\subsubsection{Norman's Emotional Design}
\begin{figure*}[hbt!]
\centering
\includegraphics[width=0.8\textwidth]{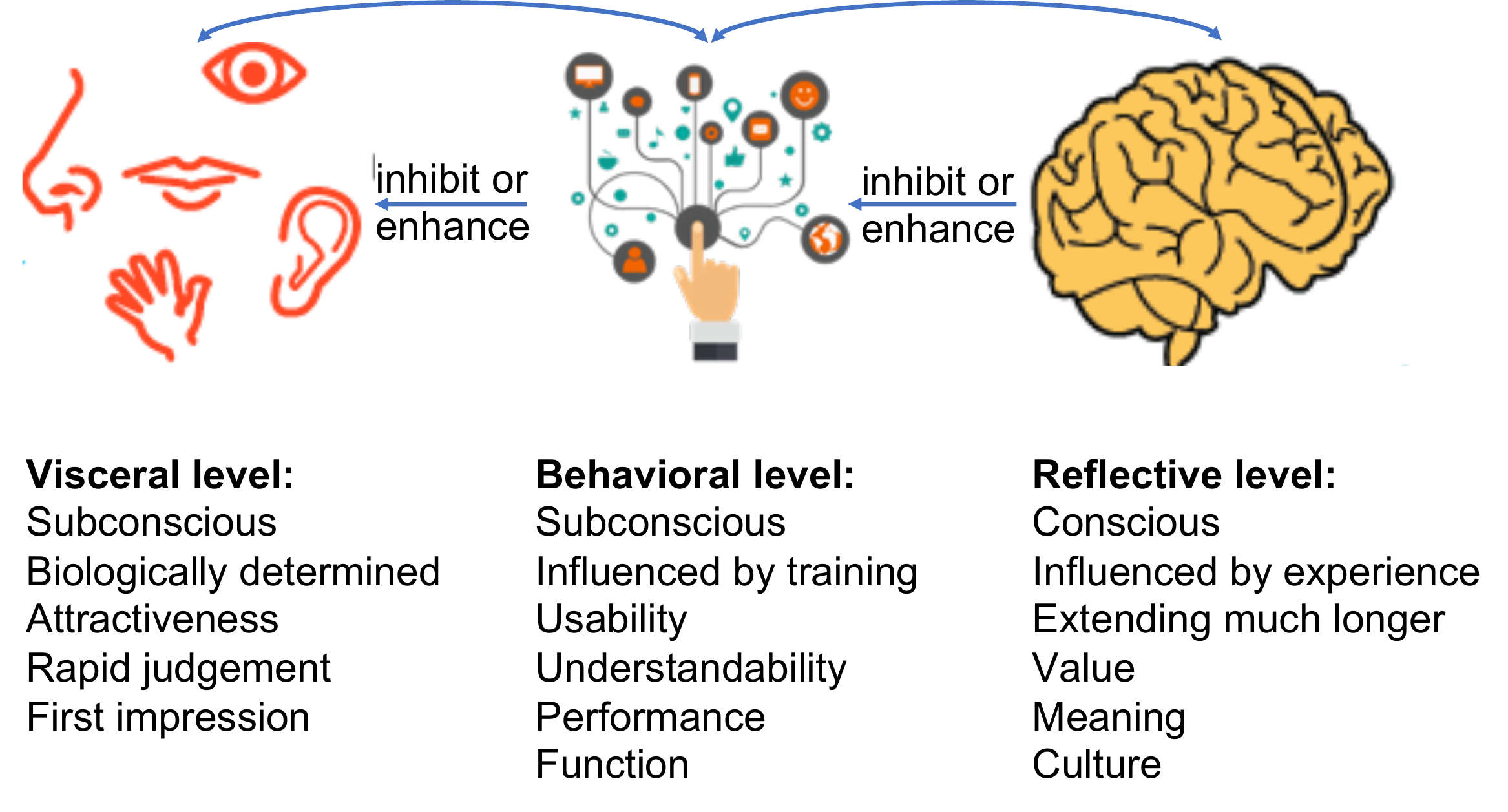}
\caption{The key features involved in Norman's emotional design} 
\label{figNormalEmotion}
\end{figure*}

The book \emph{Emotional Design: Why We Love (or Hate) Everyday Things} by Donald Norman \citeyear{norman2004emotional} described three levels of cognitive processes that give rise to emotional associations between the human user and the product, i.e., visceral, behavioral, and reflective (see Figure \ref{figNormalEmotion}). The visceral level focuses on the immediate sensory reactions to the product's physical features (e.g., look, feel, and sound), which are directly related to valenced reactions to the product (i.e., approach or avoidance). Users' visceral reactions to a product are wired in and the design principles tend to be universal. This is consistent with the baby face bias, the golden ratio rule, and the Gestalt principles mentioned above and good visceral design needs  skilled visual and industrial designers. For example, Park, Lee, and Kim \citeyear{park2011investigating} explored a new interactive touch system on a mobile touch screen by making use of the weight factor in the Laban's Effort system, and they found that it significantly improved the physical feel of the interface emotionally at the visceral level. Visceral design is prevalent in industries like automotive (e.g., Mini Cooper and Tesla vehicles), electronics (e.g., iWatch and MacBook Pro), packaging design, and so on.

The visceral level informs the behavioral level and the user subconsciously evaluates the design in terms of whether it helps complete goals with effectiveness, efficiency, and satisfaction. Behavioral design aims to improve human-product interaction, focusing on usability, performance, and function. Traditional HFE emphasizes heavily on usability and performance and in this sense, behavioral design tends to be consistent with human-centered design in that it puts user needs foremost \cite{norman2013design}. Many user research methods \cite{baxter2015understanding} in human-centered design are useful to discover user needs for good behavioral design, such as observation, ethnography, contextual inquiry, and scenario-based design. Thus, good behavioral design begins with understanding user needs, generating ideas, testing concepts, and obtaining feedback, and iteratively refines the product. For example, an autonomous system was designed for school buses using human-centered design in order to meet the needs of the parents (e.g., trust) and kids (e.g., fun, safety) at the same time \cite{ayoub2020otto}.  From affective computing point of view, systems that use behavioral (e.g., facial expressions) \cite{zhou2020fine} and physiological measures (e.g., heart rate, galvanic skin response (GSR)) \cite{zhou2011affect} to continuously monitor the human-product interaction process can potentially respond to interaction issues to improve behavioral design. For example, multiple physiological measures were used to monitoring driver states in order to improve in-vehicle system usability \cite{zhou2014augmented}.
 
With accumulation of the interaction between the user and the product, at the reflective level, the user consciously assesses the benefits, values, culture, and meaning brought by the product, which often forms emotional bonds between the user and the product. At this reflective level, the real value of the product can be way beyond the value at the visceral and behavioral levels by meeting people's affective needs, and establishing their self-image and identity in the society. A good example was described in \cite{helander2013emotional} about the user’s emotional intent or desire for vehicles and for a long time, consumption of vehicles is always more than just rational economic choices and it connects the users by aesthetic, emotional, and sensory responses to driving and symbolic relationships at both the social and cultural levels \cite{sheller2004automotive}. Unlike the previous two levels, users consciously evaluate the product at the reflective level, the real values are influenced by knowledge, experience, and culture to a great extent. For example, many special objects are associated with personal experience and memories of their own, which are often not the objects themselves, but rather the relationships and attachment to them as described in the book \emph{The Meaning of Things} by \cite{csikszentmihalyi1981meaning}. In addition, at the reflective level, users can sometimes forgive the negative experience involved at the visceral or behavioral levels. For example, long term customer experience and loyalty can often be sustained if good customer services are provided along the customer journeys by fixing defects in the initial interactions with the product, integrating multiple business functions, and creating and delivering positive customer experience \cite{lemon2016understanding}.

\subsubsection{Jordan's Four Pleasures}
Jordan \citeyear{jordan2000four} considered products as living objects that could elicit both positive and negative emotional responses and products should be designed to be useful, usable, and pleasurable. He proposed four pleasures, i.e.,  physiological, social, psychological, and ideological, to support pleasurable design. Physiological pleasure refers to pleasures generated from sensory responses, including visual, auditory, tactile, olfactory, and so on, which seem to be consistent with visceral design in Norman's emotional design. One example in vehicle related research is to make use of the odors inside the vehicle (e.g., rose compared to leather) to reduce the effect of visually induced motion sickness to improve physiological pleasure \cite{keshavarz2015visually}. Social pleasure is the enjoyment as a result of social interaction with others using the product as the medium. Direct examples are popular social media apps nowadays. Others can be the talking points involved in the social interactions, such as smart speakers, and those indicate users' specific social groups, such as Porsches for ’yuppies’ \cite{jordan1997putting}. Psychological pleasure is derived from conducting and accomplishing a task through human-product interaction, which tends to be similar to the behavioral level in Norman's emotional design. It focuses on enjoyment as a result of achieving tasks with usable products. For example, an assistance system was developed for the elderly to aid their activities in daily living and due to its proactive and case-driven characteristics, it was usable and pleasurable at the same time \cite{zhou2010case}. Ideological pleasure is related to personal aspirations and values and is derived from artistic products, such as books, music, movies, and products that embody their values. For example, some consumers were willing to buy sustainable and organic foods due to social identity and attitudes towards environmental responsibility \cite{bartels2014consumers}. Thus, products that embed such values can be popular among these consumers.

\subsubsection{Kansei Engineering}
Kansei engineering was originated in Japan as early as in the 1970s and it maps users' Kansei into product attributes in the design process using engineering methods, where Kansei is defined as \emph{the state of mind where knowledge, emotion, and passion are harmonized} \cite{nagamachi1995kansei}. The key questions in Kansei engineering are 1) how to understand Kansei accurately, 2) how to reflect and translate Kansei understanding into design elements, and 3) how to create a system and organization for Kansei-oriented design \cite{nagamachi2016innovations}. Although Kansei can be represented with different forms, adjectives are most frequently used  \cite{zhou2010hybrid} with semantic differential scales (e.g., simple - complex, spacious - narrow, boring - interesting) \cite{osgood1975cross}. 

There are three major types of Kansei engineering methods. The type I method uses a tree structure to decomposes the 0-Order Kansei concept into $n-$th order sub-concepts until these sub-concepts can be mapped to physical design elements without difficulty. The success of this method not only depends on the understanding of user Kansei, but also the decomposition of design elements that form the product. For example, a speedometer was decomposed into meter layouts, meter types and numbers, panel colors, materials, and so forth, to match user Kansei, and the contribution of each design element to specific Kansei was identified by the partial correlation coefficients based on subjective evaluation with semantic differential scales \cite{jindo1997application}. 
The type II method uses expert systems to automatically map Kansei sub-concepts to physical design elements by constructing a Kansei database, which allows the designers to understand user Kansei better. The type III method uses hybrid mapping, i.e., forward mapping from Kansei to design elements and backward mapping from design elements to Kansei. The backward mapping starts from the designers and the mapping relationships can then be revised and validated by user evaluation. For example, Zhou et al. \citeyear{zhou2010hybrid} used both $K$-optimal rule discovery for forward Kansei mapping (from Kansei to design elements) and ordinal logistic regression for backward Kansei mapping (from design elements to Kansei) to support truck cab interior design. Other methods were also proposed in order to deal with the issues involved in the previous three types, such as uncertainty of user Kansei, product element presentation (e.g., virtual reality), and effectiveness of expert systems \cite{marghani2013kansei}. For example, a deep learning method based on long short-term memory was used to extract user Kansei from online product reviews, which improved the efficiency and effectiveness of understanding user Kansei and reduced uncertainty involved in user Kansei \cite{wang2019multiple}. 

\subsubsection{Affective Computing}
Picard \citeyear{picard1997affective} coined the term \emph{affective computing} in 1997, which aims to design and develop systems that can recognize, interpret, respond to, and simulate human emotions. This is consistent with the view that emotional intelligence is one of the basic components of intelligence \cite{goleman1995emotional}. There are two major areas of research in affective computing, including 1) recognizing and responding to user emotions, i.e., affect sensing, and 2) simulating emotions in machines, affect generation, in order to enrich and facilitate interactivity between humans and machines.

First, affect sensing refers to a system that can recognize emotion by collecting data through sensors and building algorithms to recognize emotion patterns \cite{picard1997affective} based on Ekman's discrete emotion model \cite{ekman1992argument} or Russell's dimensional emotion model \cite{russell2003core}. According to the component process model of emotion \cite{scherer2005emotions}, many researchers used psychophysiological signals (e.g., GSR, electroencephalogram (EEG), heart rate), facial and vocal expressions, and/or gestures to recognize emotions. For example, GSR, facial electromyography (EMG), and EEG were used to predict emotions using a machine learning technique named rough set to recognize seven discrete emotions \cite{zhou2011affect,zhou2014emotion}. Recently, deep learning models were also used to recognize emotions, such as bi-level convolutional neural networks for fine-grained emotion recognition using Russell's dimensional emotion model \cite{zhou2020fine}. By recognizing and monitoring users' emotions, the system can respond to the users to improve learning in education \cite{wu2016review}, communications for autistic children \cite{messinger2015affective}, and video gaming \cite{guthier2016affective}, to name but a few. 

Second, many researchers simulate human emotions in social robots and virtual agents to optimize the interaction between human-robot/agent interaction. The capabilities of recognizing and expressing emotions assign characteristics to social robots and virtual agents, which can form impressions during social interactions, especially when the non-human entities are human-like, i.e., anthropomorphism \cite{eyssel2012social}. These social robots and agents can be widely applied in offices, hotels, education, personal assistants, avatars, entertainment, nursing care, therapy, and rehabilitation \cite{dautenhahn2002design,breazeal2011social,thalmann2014autonomous}. For example, a previous study showed that social robots were used as tutors or peer learners, which achieved similar cognitive and affective outcomes compared to human tutors \cite{belpaeme2018social}.  

\subsubsection{Emotional and Cognitive Design for Mass Personalization}
Mass personalization is a strategy of producing goods and services to meet \emph{individual} customers' latent needs and the surplus is positive both for customers and producers considering both the values and costs associated \cite{kumar2007mass,zhou2013affective}. Note this is different from mass customization, which aims to customize products and services for individual customers at a mass production price \cite{tseng2001mass}. The major differences are 1) mass personalization is fulfilled at the personal level, i.e., market-of-one with customer co-creation (e.g., Netflix movie recommendation), while mass customization is fulfilled for a certain market segment, i.e., market-of-few, with customer configuration (e.g., Apple computer configuration), 2) mass personalization emphasizes on high-level non-instrumental needs, including cognitive needs and emotional needs with values outperforming costs, while mass customization focuses on functional needs with near mass production efficiency, and 3) mass personalization is usually producer-initiated to delight customers with a surprise while mass customization is mostly user-initiated within the configuration defined by the producer. Furthermore, mass personalization is not personalization per se, but personalization with \emph{affordable fulfilment costs} for both customers and producers  \cite{kumar2007mass}. Many of the personalization techniques are now based on big data analytics and artificial intelligence and once the algorithms are developed, the costs associated with them tend to be minimal to provide personalized, satisfactory services for the majority of users \cite{alkurd2020big}. What mass personalization emphasizes is latent customer needs that users might not be aware of \cite{zhou2015latent}, mainly including affective and cognitive needs according to their profiles, behavioral patterns, affective and cognitive states, aesthetics preferences, and so on \cite{zhou2013affective}. We have explained affective needs above. Cognitive needs are those non-functional requirements of how products and systems are designed to accommodate human cognitive limitations \cite{zhou2011fundamentals}, which are similar to what behavioral design addresses in Norman's emotional design \cite{norman2004emotional}. Under the framework of mass personalization, we aim to integrate both affective and cognitive needs to create positive user experience throughout the product life cycle. 

\subsubsection{Summary}

\begin{table}[hbt!]
\caption{Summary of different models related to emotional design}
\label{table:1}
\begin{tabular}{|c|c|c|c|c|}
\hline
Model & Focus & Major Methods & Advantages & Limitations \\ \hline
\begin{tabular}[c]{@{}c@{}}Emotional \\ Design\end{tabular} & \begin{tabular}[c]{@{}c@{}}Visceral,\\ behavioral,\\ reflective\end{tabular} & \begin{tabular}[c]{@{}c@{}}Human-centered \\ design methods,\\ industrial design \end{tabular} & \begin{tabular}[c]{@{}c@{}}Solid theory \\ support  \\ in psychology\end{tabular} & \begin{tabular}[c]{@{}c@{}}No straightforward \\ mapping from\\ three designs\\ to specific product \\ parameters\end{tabular} \\ \hline
\begin{tabular}[c]{@{}c@{}}Four \\ Pleasures\end{tabular} & \begin{tabular}[c]{@{}c@{}}Physiological,\\  social, \\ psychological, \\ideological \\ pleasures\end{tabular} & \begin{tabular}[c]{@{}c@{}}Human-centered \\ design methods \\ (user research),\\ Pleasure analysis\end{tabular} & \begin{tabular}[c]{@{}c@{}}A framework of \\ four pleasures\\ and applicable\\ qualitative\\  research \\ methods\end{tabular} & \begin{tabular}[c]{@{}c@{}}Similar to the \\ limitations \\ of emotional \\ design\end{tabular} \\ \hline
\begin{tabular}[c]{@{}c@{}}Kansei \\ Engineering\end{tabular} & \begin{tabular}[c]{@{}c@{}}Translate \\ Kansei \\ into product \\ elements\end{tabular} & \begin{tabular}[c]{@{}c@{}}Subjective \\ evaluation,\\ expert systems\end{tabular} & \begin{tabular}[c]{@{}c@{}} Widely applied\\ in Japan \\ with successes\end{tabular} & \begin{tabular}[c]{@{}c@{}}Uncertainty of Kansei,\\ Averaged Kansei for \\ sampled participants,\\ Cultural barriers to be \\ applicable in other \\ countries\end{tabular} \\ \hline
\begin{tabular}[c]{@{}c@{}}Affective \\ Computing\end{tabular} & \begin{tabular}[c]{@{}c@{}}Emotion \\ recognition,\\ Emotion \\ generation\end{tabular} & \begin{tabular}[c]{@{}c@{}}Machine \\ learning,\\ artificial \\ intelligence\end{tabular} & \begin{tabular}[c]{@{}c@{}}Development of \\deep learning\\ models \end{tabular} & \begin{tabular}[c]{@{}c@{}}Heavily dependent on\\ models trained on \\ a specific dataset,\\privacy and moral issues\end{tabular} \\ \hline
\begin{tabular}[c]{@{}c@{}}Mass \\ Personalization\end{tabular} & \begin{tabular}[c]{@{}c@{}}Personalized\\  products for \\ individuals \\ affective and\\  cognitive needs\end{tabular} & \begin{tabular}[c]{@{}c@{}}Engineering design \\ methods, machine\\ learning models,\\human-centered \\design method \end{tabular} & \begin{tabular}[c]{@{}c@{}}Solid support in \\ engineering design\\ and human factors\\ engineering\end{tabular} & \begin{tabular}[c]{@{}c@{}}Only applicable for \\ certain products \\ and services,\\ May need big data \\ for personalization\end{tabular} \\ \hline
\end{tabular}
\end{table}

By reviewing multiple models and methods related to emotional design, we summarize them in Table \ref{table:1}. Emotional design and the four-pleasure framework are deeply rooted in human-centered design and go beyond it to include fun and pleasure. Other methods include sustainable design, participatory design, and even universal design. Thus, both can make full use of many qualitative methods in human-centered design, which are useful to incorporate affective needs in the design process. However, they do suffer some limitations, including 1) research quality can heavily depend on researchers' skills with subjectivity, 2) data analysis can be time-consuming, 3) results can be difficult to verify, and 4) there is no straightforward mapping between design elements and affective needs \cite{zhou2010hybrid,anderson2010presenting}. Kansei engineering has been widely applied in Japan with successes in different areas, such as automotive industry, cosmetics, and clothing \cite{nagamachi2016innovations}. However, the subjectivity, uncertainty, and cultural barriers associated with Kansei have restricted its applications to other countries. In addition, the designed product is often the result of the averaged Kansei of the sampled participants with perceptual preferences although other presentation methods have been proposed, such as virtual reality \cite{marghani2013kansei}. The data collection process is often time-consuming with active participation of customers and researchers \cite{wang2019multiple}. Affective computing mainly uses machine learning and artificial intelligence techniques for the machine to recognize and simulate emotions. With the development of deep learning techniques, more sophisticated and successful models have been built \cite<e.g.,>{zhou2020fine}. However, the models trained on a specific dataset can be unfair for those who are less representative (e.g., black people) in the dataset \cite{lohr2018facial}. The privacy and moral issues associated with giving the machine the capabilities to monitor and intervene users' emotional states are still under debate \cite{daily2017affective}. Mass personalization can reuse many methods in engineering design for all the steps involved and many user research methods in human-centered design can also be used, especially for affective-cognitive needs elicitation. It is built on top of mass customization and incorporates affective and cognitive needs from HFE and thus is complementary to mass customization. However, mass personalization tends to be applicable to the 'soft' characteristics of the product that are changeable and adaptable to personalize individual customers, such as those that create the experience of drinking coffee in a certain store though it is built on the 'hard' components of the product that can be configurable (i.e., mass customization), such as the coffee cups, beans, and other ingredients \cite{zhou2013affective}.   

\section{A SYSTEMATIC PROCESS FOR EMOTIONAL DESIGN} 

\begin{figure*}[hbt!]
\centering
\includegraphics[width=0.8\textwidth]{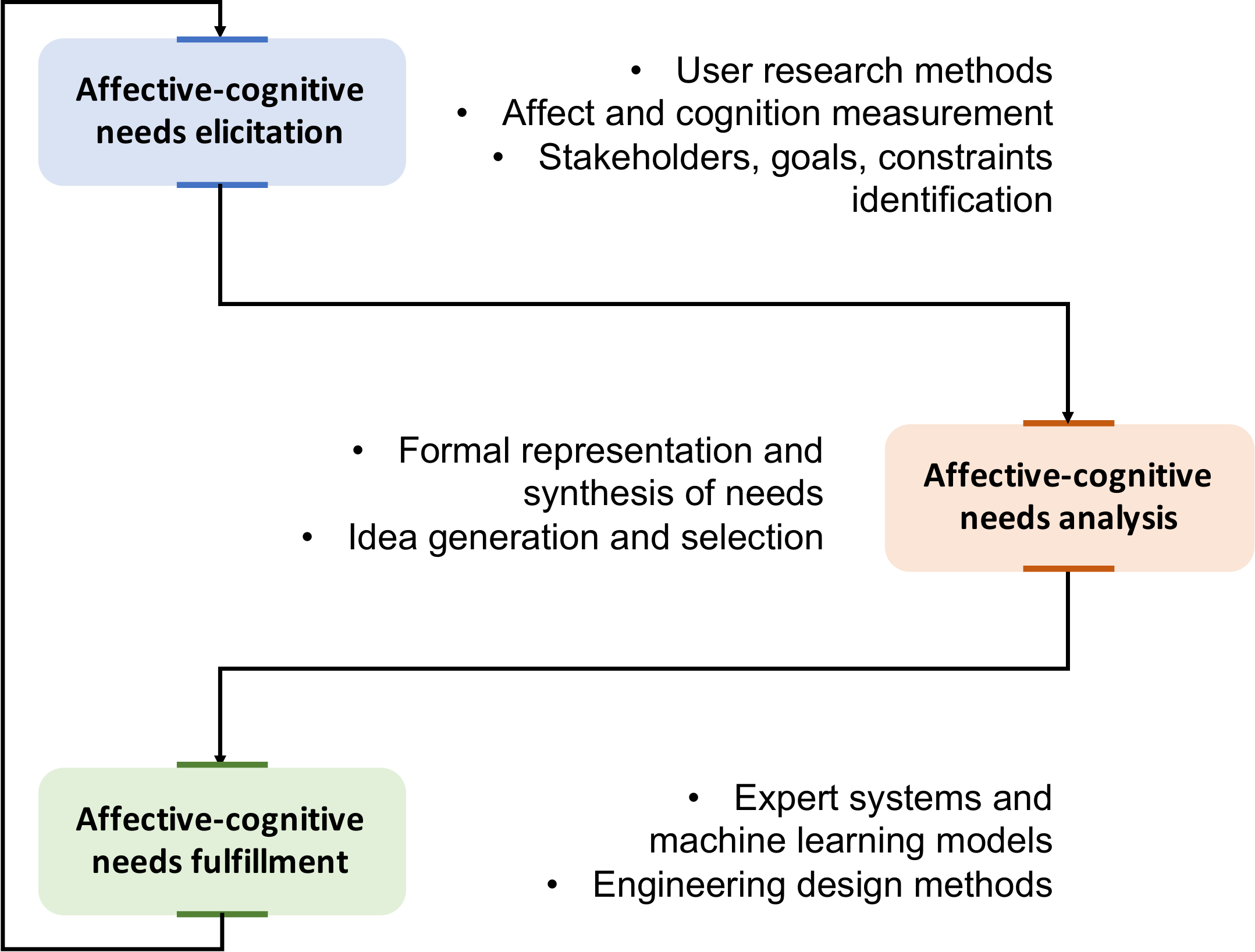}
\caption{The proposed three-step process for emotional design}
\label{figProcess}
\end{figure*}

By examining the advantages and disadvantages of different models and methods related to emotional design, we propose a three-step systematic process based on mass personalization and human-centered design to transform customers' affective and cognitive needs from the customer domain into design elements in the designer domain, including affective-cognitive needs elicitation, affective-cognitive needs analysis, and affective-cognitive needs fulfillment \cite{zhou2013affective} as shown in Figure \ref{figProcess}. The first step aims to elicit affective and cognitive needs of customers systematically, and many user research methods can be applied in this step. One of the key issues is how to measure affect and cognition constructs involved in affective and cognitive needs. At the same time, this step also identifies the involved stakeholders (e.g., customers and manufacturers), goals, use cases, and constraints. The second step aims to understand affective and cognitive needs and transform them into explicit requirements for engineers and marketers. Formal representations should be used to synthesize the needs from the first step, concepts should be generated and selected based on the priorities of customer needs. The last step aims to identify the mapping relationships between the requirements and product specifications with an iterative process of prototype testing. This three-step process itself should also be iterative to refine the product. At the same time, many machine learning models and expert systems involved in affective computing and Kansei engineering can also be used to support emotional design. The related work is reviewed below.

\subsection{Affective-cognitive Needs Elicitation and Measurement}
\subsubsection{User Research for Needs Elicitation}
Many user research methods in human-centered design have been proposed for affective and cognitive needs elicitation \cite{baxter2015understanding}. For example, in order to understand the cognitive needs of healthcare workers when designing medical software, Johnson and Turley \citeyear{johnson2006significance} used a think-aloud protocol. A diary study (with introductory, mid-study, and final interview with each participant, spaced 7 days apart) was conducted over two weeks to understand the informational needs of mobile phones \cite{sohn2008diary}. Observation was used in public transportation, such as trains, in order to understand user needs to support none-driving related tasks in automated vehicles \cite{pfleging2016investigating}.  Contextual inquiry was used to gain a deeper understanding of how drivers interact with vehicles' infotainment systems to create positive driver-vehicle interactive experience \cite{gellatly2010journey}.  For more examples, please refer to \cite{baxter2015understanding}.

For affective needs elicitation, Ng and Khong \citeyear{ng2014review} reviewed various methods for affective human-centered design for video games and proposed two types of methods, including user-feedback methods (e.g., focus group, survey, interviews, usability testing methods) and non-intrusive methods (e.g., observation on facial and vocal expressions, physiological sensors). Then, Ng, Khong, and Nathan \citeyear{ng2018evaluating} applied interviews as a user-feedback method and observation as a non-intrusive method to affective video game design, where the interviews were used to measure subjective feelings while the observation was used to understand participants’ emotional responses during their game playing. Many researchers in Kansei engineering applied surveys, questionnaires, and focus groups to collect Kansei from users \cite{wang2019multiple}. For example, Quan, Li, and Hu \citeyear{quan2018product} used questionnaires to elicit Kansei by reviewing clothes images from both designers and consumers. Kwong, Jiang, and Luo \citeyear{kwong2016ai} utilized conjoint and lead user surveys to understand customer Kansei of electric irons.  Akay and Kurt \citeyear{akay2009neuro} interviewed users and surveyed magazines to understand customer Kansei of mobile phones. The sample sizes in these Kansei studies were relatively small ($<20$) and in order to reduce the possible subjective biases \cite{pryzant2020automatically}, online product reviews can be readily collected from websites (e.g., Amazon.com) in large quantities. For example, a large amount of review data for Kindle tablets and Amazon product ecosystems were crawled from Amazon to understand reviewers' emotional responses and satisfaction \cite{zhou2015latent,ayoub2019analyzing,zhou2020machine}. Human agents or avatars are also used to elicit emotional responses for interactive interfaces. For example, an affective avatar was designed based on a human–avatar taxonomy to address social communication disorders \cite{johnson2018assessing}. Wizard-of-oz methods were used in automated driving by hiding the drivers to elicit emotional responses and natural behaviors of the passengers \cite{ayoub2020otto} and pedestrians \cite{currano2018vamos}. Methods that can elicit both affective and cognitive needs at the same time are reported, too. For example, Coursaris and van Osch \citeyear{coursaris2016cognitive} proposed a cognitive-affective model of perceived user satisfaction and used experiment design by manipulating colors in website design to understand participants' perceived cognitive (effectiveness and efficiency) and emotional (aesthetics and playfulness) responses.   

\subsubsection{Affect and Cognition Measurement}

\textbf{Subjective Measures:} These methods are probably the most frequently used ones to measure affect (e.g., emotional responses and feelings) and cognition (e.g., cognitive workload) with efficiency. For example, in affective computing, in order to train a model that can recognize emotions, the ground truth used as labels in training is often produced by subjective self-reports. For example, Zhou et al. \citeyear{zhou2011affect,zhou2014emotion} used participants' self-reported emotional responses to static images and sound clips as labels to train machine learning models. One of the possible issues is the forced-choice method among a list of discrete emotions leading to a relative judgement, which may not necessarily reflect the participant's real emotional responses \cite{russell1993forced}. In order to obtain a reliable set of labels, crowd-sourcing labeling with multiple workers can be useful. For example, Barsoum et al. \citeyear{barsoum2016training} used multiple crowd workers to label each facial expression image and found the agreement increased from less than 40\% with 3 workers to over 90\% with 9 workers. For dimensional emotion recognition, the reliability is often measured among individual raters, such as AffectNet  \cite{mollahosseini2017affectnet}.

In Kansei engineering, researchers often applied semantic differential scales with participants' self-reported measures \cite{nagamachi2016innovations}. For example, Lu and Petiot \citeyear{lu2014affective} applied semantic differential scales in designing eyeglasses. Other useful methods include self-assessment manikin on valence, arousal,and dominance \cite{bradley1994measuring}, the product emotion measurement instrument (PrEmo) using animations of cartoon characters with a small number of basic emotions \cite{desmet2003values}, and the experience sampling method to collect daily experience over a longer period of time \cite{larson2014experience}. For instance, the self-assessment manikin instrument was applied to measure valence and arousal and their influence on the interaction between drivers and automated vehicles \cite{du2020examining} and the experience sampling method was used to identify the antecedents of daily positive emotions \cite{goetz2010antecedents}. 

In order to measure cognitive constructs, many survey-based tools have been developed. The most frequently used survey tools for measuring cognitive workload are probably the NASA Task Load Index \cite{hart2006nasa}, the Workload Profile \cite{tsang1996diagnosticity}, and the Subjective Workload Assessment Technique \cite{reid1988subjective}. For example, Ayoub and Zhou \citeyear{ayoub2020investigating} used the NASA Task Load Index to measure cognitive workload of automated vehicle interfaces in the context of lane changing events. Rubio et al. \citeyear{rubio2004evaluation} compared these three tools and found all of them had good validity, but the Workload Profile had better sensitivity and diagnostic powers. 

\textbf{Objective Measures:} Subjective measures are easy to implement. However, they tend to be susceptible to subjective biases \cite{pryzant2020automatically}. Objective measures, such as behavioral and physiological measures, are less susceptible to voluntary control. The typical behavioral measures include facial and vocal expressions, poses, and gestures while physiological measures include eye tracking data, GSR, heart and respiration activity, EMG, EEG, and so on. For example, eye tracking data was used to measure attention, GSR was used to measure arousal, and heart rate and heart rate variability were used to measure cognitive workload in the interaction between the driver and automated vehicles \cite{du2020examining}. Facial expressions were used to understand participants' dimensional emotional states \cite{zhou2020fine} and trust in human-automation interaction \cite{neubauer2020analysis}. Due to the fact that one measure is not able to reliably measure emotion or cognition, many researchers often use multiple measures together. For example, both facial expressions and head poses were used to measure emotions and understand human interaction by visualizing depth information \cite{kalliatakis2017conceiving}. GSR, respiration rate, facial EMG, and EEG were used to measure participants' emotional responses to visual and auditory stimuli \cite{zhou2011affect,zhou2014emotion}. 
Koelstra et al. \citeyear{koelstra2011deap} used 32 channels of EEG data, electrooculography, zygomaticus major EMG, trapezius EMG, GSR, respiration, plethysmograph, and peripheral skin temperature to measure participants' emotional responses. Capitalizing on this dataset, many researchers applied machine learning models to recognize emotions, \cite<e.g.,>{piho2018mutual,cui2020eeg}.

From the behavioral point of view, another important tool in understanding the cognitive demands, thought processes, knowledge, and goals is cognitive task analysis, which combines the features of the work domain and the cognitive demands imposed on the user \cite{schraagen2000state}. There are many techniques (including subjective methods) used in cognitive task analysis, such as critical incident/decision analysis, cognitive field observation, hierarchical task analysis, sequence analysis, knowledge audit, and so on \cite{crandall2006working}. For example, main display patterns and themes were identified using cognitive task analysis to support software design \cite{pfautz2006using}.  Zanesco \citeyear{zanesco2020quantifying} applied sequence analysis to understand the dynamic thought process using time series data across different cognitive tasks. 

\subsection{Affective-cognitive Needs Analysis}
Some of the elicitation methods mentioned above have their analysis components, such as cognitive task analysis \cite{crandall2006working}. Since affect and cognition tend to be qualitative in nature, many of the methods for affective-cognitive needs analysis are qualitative methods and were used in human-centered design, such as grounded theory \cite{strauss1994grounded} and affinity diagram \cite{spool2004kj}. However, qualitative methods can be time-consuming when the data amount is large. With the development of big data and machine learning techniques, many efficient and quantitative methods are proposed. 

\subsubsection{Qualitative Methods}
Qualitative methods can potentially produce different types of representation of needs, such as profiles, patterns and themes, importance and priorities, concepts and classifications, and so on. Among many, personas are often created as profiles of archetype users in order to analyze users' needs by scrutinizing their goals, needs, wants, and pains, based on which different scenarios can be created for activities of empathetic role-play \cite{pruitt2010persona}. For example, personas were used to analyze emotional needs for automated public transportation services in a multi-stakeholder context \cite{kong2018personas}. 
Affinity diagram, developed by Jiro Kawakita and also named the KJ method, is widely used for customer needs analysis in terms of identifying themes and patterns and assigning importance and priority \cite{spool2004kj}. For example, over 800 Kansei words were grouped into 43 clusters using an affinity diagram for web design in four different steps, including initial study, exploratory study, KJ method, and confirmatory study \cite{lokman2010kansei}. Ayoub et al. \citeyear{ayoub2020otto} also used an affinity diagram to prioritize the emotional needs of parents and kids based on personas for designing an automated school bus. Another important method is grounded theory  \cite{strauss1994grounded} and it aims construct theories and identify patterns and themes using specific coding schemes of the data systematically, especially text data generated in the elicitation process. For instance, Brown and Cairns \citeyear{brown2004grounded} used grounded theory to categorize immersion into engagement, engrossment, and total immersion for game design, and such grouping was able to be applied to software design. Zhou, Yang, and Zhang \citeyear{zhou2020takeover} applied grounded theory to code comments on YouTube videos of automated driving and identified major human factors issues of automated vehicles.
In addition, both task analysis and cognitive task analysis can be used to identify themes and concepts to support decision making. For example, a task-based needs analysis was proposed to obtain insights from task selection, task discourse analysis, task difficulty, and task sequencing for designing foreign language instructions \cite{malicka2019needs}. The critical decision method as an approach to cognitive task analysis was used to identify a list of critical cues and judgements, such as action, knowledge, appraisal, and anticipation in training \cite{hoffman1998use}. 

Graphic methods are also useful in affective-cognitive analysis, such as mind maps, concept maps, and cognitive maps. A mind map can be used to visually represent hierarchical relationships among pieces of information, usually with one focus \cite{hopper2012practicing}. For example, individual mind maps from each participant were used to represent Kansei words and then an overall mind map was developed by aggregating individual ones \cite{huang2014product}. Unlike mind maps, a concept map is defined as a relationship diagram using labeled arrows (e.g., "consist of", "give rise to") to connect different concepts in a hierarchical structure \cite{novak2006theory}. For example, concept maps were used to represent knowledge of retired NASA engineers to help train novices \cite{coffey2003graphical}. Cognitive maps use causal links to represent concepts and it was used to represent the decision-making process of different team members in new product design \cite{carbonara2006cognitive}. Fuzzy cognitive maps incorporate fuzziness involved in the relationships between concepts and were used to capture the causal reasoning process in geographic information system design \cite{liu1999contextual}. 

When affective and cognitive needs are synthesized, potential solutions can be generated to satisfy these needs. In human-centered design, idea generation or brainstorming aims to generate as many ideas as possible in the first place and then systematic analysis can be done to select the optimal candidates. Sketching is widely used in idea generation because it is quick, inexpensive, disposable, plentiful, with distinct gestures and minimal details \cite{buxton2010sketching}. Scribble sketching can rapidly sketch the idea anytime, anywhere and is used not only to generate ideas but also collect existing ideas while 10puls10 aims to generate 10 or more ideas and then select the most promising one to generate 10 detailed variations \cite{greenberg2011sketching}. While these techniques are widely used in idea generation, many researchers tend to develop new sketching tools, especially digital ones. For example, Spatial Sketch was developed as a 3D sketch application to combine physical movement and object fabrication in the real world using cut planar materials \cite{willis2010spatial}. DataToon included elements of comics to create data-driven storyboards that blended analysis and presentation with pen and touch interactions \cite{kim2019datatoon}. 
In order to generate a large number of ideas, crowdsourcing \cite{majchrzak2013towards} has been widely used. For example, Schuurman et al. \citeyear{schuurman2012smart} explored  crowdsourcing for idea generation and selection for smart city innovation. In order to deal with a large volume of the ideas produced by crowd sourcing, Hoornaert et al. \citeyear{hoornaert2017identifying} identified three sources to select ideas, including the content, the contributor, and the crowd's feedback on the idea. Faste et al. \citeyear{faste2013brainstorm} proposed new ideas using digital collaborative ideation, including chainstorming (i.e., passing ideas along the communication chain), cheatstorming (i.e., brainstorming without generating ideas), and tweetstorming (i.e., a digital chainstorming that used cheatstorming) and found that brainstorming was not only pooling existing ideas but also involving the sharing and interpretation of  concepts in unintended and unanticipated ways.

\subsubsection{Quantitative Methods}
One of the limitations of the qualitative methods mentioned above is that the data analysis process tends to be laborious and subjective. For example, qualitative persona methods tend to be time-consuming if the designers have to examine a large number of participants. In order to overcome this issue, Hence, Zhang, Brown, and Shankar \citeyear{zhang2016data} proposed a quantitative method by making use of the click streams from 2400 users and a hierarchical clustering model to identify typical personas in order to improve user experience. The grounded theory and affinity diagram also suffer from similar issues for analyzing a large amount of text data. In order to automate this process, recently, natural language processing techniques have achieved great successes in various tasks, especially for those with deep learning models \cite{devlin2018bert}. Zhou et al. \citeyear{zhou2020machine} used latent Dirichlet allocation to identify the topics from over 90,000 online reviews on Amazon product ecosystems and sentiment analysis was used to automatically predict their sentiment polarity and intensity. Wang et al. \citeyear{wang2019multiple} proposed a heuristic deep learning model to automatically generate multiple Kansei pairs by mining a large number of online product reviews. 

Other quantitative methods aim to group, quantify, or prioritize affective and cognitive needs with statistical criteria, which can potentially overcome subjective biases \cite{pryzant2020automatically} involved in the qualitative methods. For example, principal component analysis was used to identify the major Kansei concepts among the collected data \cite{barnes2009decision} and a Kansei clustering method was proposed with design structure matrices, where partial correlation coefficients were used as the distances between Kansei adjectives \cite{huang2012kansei}. Similarly, a combination of design structure matrices and genetic algorithms was used to identify the connections between Kansei for optimal clusters \cite{yang2016consumers}. Conjoint analysis was used to measure utilities of product profiles linked to the affective needs of truck cabs based on ratings on Likert scales \cite{jiao2007analytical}. 

Affective and cognitive needs are also associated with different levels of uncertainties and ambiguities due to their qualitative nature. Researchers also proposed quantitative methods to deal with this issue. For example, due to the individual differences among participants, Kansei individuality was modeled using fuzzy set theory \cite{nakamori2004modeling}. Grey relationship degree analysis was used to identify the priority of Kansei adjectives \cite{kang2020combining}. Fuzzy product rules based on rough set were used in order to deal with the uncertainty, complexity, and dynamics associated with user experience modeling and quantification, which included both affective needs and cognitive needs \cite{zhou2011user}. Zhai et al. \citeyear{zhai2009dominance, zhai2009rough} made use of the rough numbers in rough set to model the uncertainties involved in affective and cognitive needs to produce reliable priorities. Li et al. \citeyear{li2017rule} applied evidence theory's reliability indices (e.g., support and confidence) to the rules generated by rough set using neural networks to improve precision in Kansei knowledge. Su et al. \citeyear{su2020novel} used convolutional neural networks to identify the importance of different Kansei attributes to overcome subjective biases \cite{pryzant2020automatically}.

As a quantitative technique, deep learning has been applied to automatically generate design concepts. For example, Raina, McComb and Cagan \citeyear{raina2019learning} used a deep convolutional autoencoder to imitate human designers to generate high-level semantic information from image designs without any objectives. Recently, a more powerful generative deep learning model, i.e., generative adversarial nets \cite{goodfellow2014generative}, has been proposed and it has better capabilities to extract key information contained in the design space to generate new designs and requires minimal input from the designer. For example, Shu et al. \citeyear{shu20203d} trained a generative adversarial network to generate 3D aircraft models and after three iterations of the training-evaluation process, the produced design had statistically significant improvement based on evaluation in a simulated environment. Chen et al. \citeyear{chen2019artificial} proposed a semantic ideation network and a visual concept combined model based on generative adversarial network and their model was able to generate cross-domain concepts efficiently with quantity and novelty. The selection process was conducted by domain experts.

\subsection{Affective-cognitive Needs Fulfilment}
The fulfillment step aims to create mapping relationships between affective and cognitive needs of customers and product specification in the form of design features or elements. Both traditional engineering methods and machine learning models are widely used though many engineering design methods apply machine learning models, too.

\subsubsection{Quality Function Deployment} 
One of the most frequently used engineering design methods to translate customer needs, especially functional needs, to product specifications is probably quality function deployment  \cite{prasad1998review}. Quality function deployment is a method that first transforms qualitative customer needs into quantitative parameters, then deploys the functions to form  product quality, and then translates product quality into design elements, and finally to specific manufacturing processes \cite{akao1994development}. For example, Jin et al. \citeyear{jin2009development} used quality function deployment to develop evaluation models for overall emotional factors, detailed emotional factors, usability factors, and physical design specifications in three sequential processes, based on which the emotional factors affecting physical design specifications were generated. In order to better deal with the uncertainty involved in affective and cognitive needs, fuzzy and rough quality function deployment was also used. For example, Kang et al. \citeyear{kang2018integrating} integrated the evaluation grid method with fuzzy or rough quality function deployment to build relationships between affective needs and design elements, where a fuzzy analytic hierarchy process was integrated with quality function deployment to prioritize the affective needs involved. Later, Kang \citeyear{kang2020aesthetic} applied both fuzzy quality function deployment and rough set theory to develop relationships between aesthetic product elements and customer satisfaction. Similarly, Zhai et al. \citeyear{zhai2008rough} combined quality function deployment with rough set, where rough numbers were used to help deal with subjective variables involved in affective needs and design parameters.

\subsubsection{Machine Learning Methods} 
Regression models are among the first to construct the relationships between the customer needs and product design elements. For instance, a general linear regression model was used to connect design specifications and participants' emotional responses to website design \cite{kim2003designing}. Likewise, a multiple regression model was used to link usability factors to design elements \cite{han2000evaluation}. However, linear relationships might not well represent the relationships between customer needs and design specifications \cite{zhou2010hybrid}. 

One of the frequently used nonlinear methods is association rule mining that can directly link needs and design elements using if-then rules \cite{jiao2006kansei}. Using goodness evaluation of the rules, helpful rules can be identified \cite{zhou2010hybrid}. The uncertainty involved in customer needs can be mitigated using fuzzy set theory as mentioned above. Kwong, Jiang, and Luo \citeyear{kwong2016ai} used chaos-based fuzzy regression to understand both the concerns and satisfactions of design, engineering, and marketing issues in Kansei engineering. By combing if-then rules and fuzzy set theory, Akay and Kurt \citeyear{akay2009neuro} proposed a neuro-fuzzy if-then rules to identify the relationships between physical form design elements and customers' affective responses for mobile phone design.  However, the association rules identified can still be spurious if they are found just by co-occurrences. Another method often used is neural networks and the powerful nonlinear modeling capability can help identify the relationships between affective-cognitive needs and design attributes. For example, Hsiao and Huang  \citeyear{hsiao2002neural} used neural networks to build the connection between product form parameters and Kansei adjectives. Kang \citeyear{kang2020combining} applied neural networks as a mapping function to identify the important Kansei factors and product design elements for vehicle booth design. Many researchers made use of the advantages of multiple machine learning models for the fulfillment task. Quan, Li, and Hu \citeyear{quan2018product} proposed a deep transfer learning model to generate new product models by reconstructing and merging color and pattern features for clothes, and then used a neural network model to identify the relationships between product elements and Kansei. Wang \citeyear{wang2011hybrid} proposed a combined approach of grey system theory and support vector regression to capture the bidirectional relationships between customers' affective needs and design elements. Recently, deep learning models have also been applied in this fulfillment task and compared to traditional machine learning models, deep learning models are more successful. For example, Wang et al. \citeyear{wang2018classification} applied two deep learning techniques, named Deep Belief Network and Restricted Boltzmann machines, to classify multiple affective attributes of customer reviews and compared to traditional learning models they used, i.e., support vector machine and softmax regression, the accuracy of the deep learning models was 50\% higher.   

\section{CHALLENGES AND FUTURE DIRECTIONS}
Upon reviewing work published in the past related to emotional design, we speculate recent trends and possible future directions below.
 
\subsection{Measuring Emotion and Cognition in Naturalistic Setting}
Emotion is dynamic and short-lived and many studies \cite<e.g.,>{du2020examining} presented above measured emotion in controlled laboratories and some were limited to a small number of basic emotions  \cite<e.g.,>{zhou2011affect,zhou2014emotion}. However, in many scenarios, emotion recognition should be conducted in naturalistic settings and the number of emotions in different applications need to go beyond the six basic emotions \cite{zhou2020fine}. Regardless of the good performance of emotion recognition from the reported studies in laboratory conditions, applications using emotion recognition in naturalistic settings still remain an open challenge, such as in the context of learning, driving, entertainment, and robotics \cite{avots2019audiovisual}. For example, in the Affective Behavior Analysis in-the-wild 2020 Competition, it was reported that the best average concordance correlation coefficient for valence and arousal recognition was only 0.447, and the best weighted performance between accuracy and f1-score for seven basic emotion and eight action unit detection was only 0.509 and 0.607, respectively \cite{kollias2020analysing}. 

Another challenge is the ambiguities and uncertainties involved in subjective human emotion and cognition, as evidenced in Kansei engineering. Furthermore, many studies only recruited a small number of participants and the sampled Kansei might be biased for the targeted user groups. For example, only four participants were reported in \cite{jiang2015rough} and five participants were reported in \cite{kwong2016ai}. Due to the time-consuming data collection process while customer needs might change from time to time rapidly with an increasing number of new products in the market, it is necessary to develop an efficient method to collect data from a large number of participants continuously \cite{wang2019multiple}. 

One of the possible solutions to deal with such challenges is to make use of the technologies of the internet of things with the convergence of deep learning techniques, commodity sensors, and embedded systems. Such technologies have contributed a surge in data traffic, making real time measuring of user states, including emotions and cognitive states a possibility with advanced deep learning models and computational resources \cite{gubbi2013internet}. First, deep learning techniques have achieved great successes in many applications, such as computer vision \cite{hassaballah2020deep}, natural language processing \cite{devlin2018bert}, and emotional and cognitive states recognition \cite{zhou2020fine}. Second, the development of the internet of things offers enough training data to improve the performance of deep learning models \cite{chan2020affective}. For example, wearable sensors (e.g., smart watches) and other commodity sensors can be readily used to collect various data about the user (physiological and behavioral data) on the fly over wireless networks \cite{alkurd2020big}. In addition, human-computer integration is emerging, in which computational and human systems can be interwoven closely in a wider social-technical system \cite{mueller2020next}. Furthermore, user-generated data on websites, such as online product reviews on Amazon.com, can be utilized to understand and update customer needs more efficiently and effectively with a large number of users on a daily basis \cite{li2018dynamic, zhou2020machine}. Third, high-performance computing resources, such as graphic processing units and tensor processing units, allow the training of large-scale deep learning models for big data possible \cite{zhang2018survey}. Under such circumstances, smart and personalized services based on artificial intelligence that can respond to user feedback immediately might be the cardinal competitive advantage for all service providers \cite{alkurd2020big}. At the same time, the contextual data should also be utilized in order to optimize and personalize the interaction process between humans and the system \cite{zhou2011fundamentals}. 

\subsection{Integration of Affect and Cognition}
Affect and cognition have been treated as independent entities \cite{zajonc1980feeling}. However, studies have shown that affect and cognition are highly interrelated and should be integrated \cite{jiao2017decision}. For example, Kahneman \citeyear{kahneman2003perspective} pointed out that there were two systems for decision making, i.e., the analytic system (i.e., the cognitive system) and the experiential system (i.e., the affective system). The analytic system deliberately uses cognitive processes to make reasonable decisions while the experiential system operates outside of conscious thoughts and utilizes emotion-related associations, past experience and intuitions for reactive decision making.  The behavioral level in Norman's emotional design directly points out
that understandability and usability are the two key factors in contributing to positive emotional responses \cite{norman2004emotional}. The appraisal theory emphasizes the cognitive component of emotions, which can be used to distinguish different emotions \cite{ortony1990cognitive}. Ahn and Picard \citeyear{ahn2005affective} also proposed an integrated framework of affect and cognition for decision and learning. Such integrated perspectives were also emphasized in our previous work for user experience modeling and design \cite{zhou2011fundamentals,zhou2013affective,zhou2014prospect}. 

Based on the view of cognitive and affective systems when we process information, Norman's emotional design gives us good guidelines and implications at three different levels as well as the interactions between affect and cognition. For example, positive affect allows us to think more broadly and creatively, supporting idea generation \cite{isen1987positive} while negative affect narrows one's cognitive scope \cite{rathunde2000broadening}. However, a recent study suggested that motivational intensity (i.e., how strongly one was compelled to approach or avoid something) that influenced one's cognitive scope rather than emotional valence \cite{harmon2013does}. In their study, they found that 1) amusement (a low motivational emotion) elicited by watching a cat video broadened the participants' attentional focus while desire (a high motivational emotion) elicited by watching a delicious-looking dessert video narrowed their attentional focus and 2) sadness (a low motivational emotion) broadened their attentional focus while disgust (a high motivational emotion) narrowed their attentional focus. 
Therefore, despite the common consensus on integrated affect and cognition, the interaction between affect and cognition tends to be less clear \cite{storbeck2007interdependence} and it is challenging to develop analytic models to integrate affective and cognitive needs, to measure subjective experience, and to extract the mapping relationships between integrated affective and cognitive needs and design elements \cite{jiao2017decision}. More research in this aspect is called for to support emotional design.

\subsection{Emotional Design for Product Ecosystems}
Many companies now focus more on the overall experience of their product ecosystems rather than one single individual product in order to improve their competitiveness. A product ecosystem has a focal product at the center during the human-product interaction process, with numerous other peripheral supporting products and services to deliver an entire experience so that other more disjointed offerings cannot compete \cite{zhou2011fundamentals}. Good examples include the Apple product ecosystem and the Amazon product ecosystem. From the experiential point of view, the roles of affect and cognition as an antecedent, a consequence, and a mediator of human-product-ambience interaction within a product ecosystem over time form one's total user experience \cite{hassenzahl2006user,zhou2011fundamentals}. First, within a product ecosystem, when one is interacting with a product, that product becomes the focal product and others become supporting products, named as ambience. Without the supporting role of other ambient products, the interaction with the focal product might be isolated in a narrower scope. For example, Levin \citeyear{levin2014designing} emphasized the interaction between multiple devices, such as smart phones, computers, tablets, and TVs to create user experience with an ecosystem approach. Gawer and Cusumano \citeyear{gawer2014industry} pointed out that within a company's platform, numerous derivative products and services should be designed within a common ecosystem perspective to support innovation and user experience and within an industry platform, complementary products, services, and technologies should be designed and developed within a business ecosystem to promote innovation. Second, the role of affect and cognition should be expanded from just the consequence of design. Traditional affective computing emphasizes to respond retrospectively to users' emotions during the interaction process to aid frustrated users \cite{picard2002computers}. Within a product ecosystem, the system should proactively predict users' emotions and examine the roles of affect and cognition as the antecedent and mediator factors as well. Third, we emphasize not only the temporary interaction between the user and the ecosystem but also over a longer period of time to form a totality of user experience. This echoes Norman's reflective design to some extent. With the complexity and dynamics of user experience over a longer period of time, emotional designers need to consider all relevant factors in the product ecosystem. This is consistent with the perspective of measuring emotion in the naturalistic setting and advanced technologies of the internet of things with convergences of deep learning, big data, and wireless networks could potentially help.

\section{CONCLUSIONS}
Emotional design has been well recognized in the domain of human factors and ergonomics. In this chapter, we reviewed related models and methods of emotional design.  We are motivated to encourage emotional designers to take multiple perspectives when examining these models and methods. Then we proposed a systematic process for emotional design, including affective-cognitive needs elicitation, affective-cognitive needs analysis, and affective-cognitive needs fulfillment to support emotional design. Within each step, we provided an updated review of the representative methods to support and offer further guidance on emotional design. We hope researchers and industrial practitioners can take a systematic approach to consider each step in the framework with care. Finally, the speculations on the challenges and future directions can potentially help researchers across different fields to further advance emotional design. 
\bibliographystyle{apacite}
\bibliography{sample}

\end{document}